\documentstyle[twoside]{article}

\catcode`\@=11
\long\def\@makefntext#1{
\protect\noindent \hbox to 3.2pt {\hskip-.9pt  
$^{{\eightrm\@thefnmark}}$\hfil}#1\hfill}               %CAN BE USED 

\def\thefootnote{\fnsymbol{footnote}}
\def\@makefnmark{\hbox to 0pt{$^{\@thefnmark}$\hss}}    %ORIGINAL 
        
\def\ps@myheadings{\let\@mkboth\@gobbletwo
\def\@oddhead{\hbox{}
\rightmark\hfil\eightrm\thepage}   
\def\@oddfoot{}\def\@evenhead{\eightrm\thepage\hfil
\leftmark\hbox{}}\def\@evenfoot{}
\def\sectionmark##1{}\def\subsectionmark##1{}}

\oddsidemargin=\evensidemargin
\renewcommand{\thefootnote}{\fnsymbol{footnote}}

\newcounter{sectionc}\newcounter{subsectionc}\newcounter{subsubsectionc}
\renewcommand{\section}[1] {\vspace{12pt}\addtocounter{sectionc}{1} 
\setcounter{subsectionc}{0}\setcounter{subsubsectionc}{0}\noindent 
        {\tenbf\thesectionc. #1}\par\vspace{5pt}}
\renewcommand{\subsection}[1] {\vspace{12pt}\addtocounter{subsectionc}{1} 
        \setcounter{subsubsectionc}{0}\noindent 
        {\bf\thesectionc.\thesubsectionc. {\kern1pt \bfit #1}}\par\vspace{5pt}}
\renewcommand{\subsubsection}[1] {\vspace{12pt}\addtocounter{subsubsectionc}{1}
        \noindent{\tenrm\thesectionc.\thesubsectionc.\thesubsubsectionc.
        {\kern1pt \tenit #1}}\par\vspace{5pt}}

\newcounter{appendixc}
\newcounter{subappendixc}[appendixc]
\newcounter{subsubappendixc}[subappendixc]
\renewcommand{\thesubappendixc}{\Alph{appendixc}.\arabic{subappendixc}}
\renewcommand{\thesubsubappendixc}
        {\Alph{appendixc}.\arabic{subappendixc}.\arabic{subsubappendixc}}

\renewcommand{\appendix}[1] {\vspace{12pt}
        \refstepcounter{appendixc}
        \setcounter{figure}{0}
        \setcounter{table}{0}
        \setcounter{lemma}{0}
        \setcounter{theorem}{0}
        \setcounter{corollary}{0}
        \setcounter{definition}{0}
        \setcounter{equation}{0}
        \renewcommand{\thefigure}{\Alph{appendixc}.\arabic{figure}}
        \renewcommand{\thetable}{\Alph{appendixc}.\arabic{table}}
        \renewcommand{\theappendixc}{\Alph{appendixc}}
        \renewcommand{\thelemma}{\Alph{appendixc}.\arabic{lemma}}
        \renewcommand{\thetheorem}{\Alph{appendixc}.\arabic{theorem}}
        \renewcommand{\thedefinition}{\Alph{appendixc}.\arabic{definition}}
        \renewcommand{\thecorollary}{\Alph{appendixc}.\arabic{corollary}}
        \renewcommand{\theequation}{\Alph{appendixc}.\arabic{equation}}
        \noindent{\tenbf Appendix \theappendixc #1}\par\vspace{5pt}}
\newcommand{\subappendix}[1] {\vspace{12pt}
        \refstepcounter{subappendixc}
        \noindent{\bf Appendix \thesubappendixc. {\kern1pt \bfit #1}}
        \par\vspace{5pt}}
\newcommand{\subsubappendix}[1] {\vspace{12pt}
        \refstepcounter{subsubappendixc}
        \noindent{\rm Appendix \thesubsubappendixc. {\kern1pt \tenit #1}}
        \par\vspace{5pt}}

\newcommand{\textlineskip}{\baselineskip=13pt}
\newcommand{\smalllineskip}{\baselineskip=10pt}

\def\eightcirc{
\begin{picture}(0,0)
\put(4.4,1.8){\circle{6.5}}
\end{picture}}
\def\eightcopyright{\eightcirc\kern2.7pt\hbox{\eightrm c}} 

\newcommand{\copyrightheading}[1]
        {\vspace*{-2.5cm}\smalllineskip{\flushleft
        {\footnotesize International Journal of Modern Physics A, #1}\\
        {\footnotesize $\eightcopyright$\, World Scientific Publishing
         Company}\\
         }}

\newcommand{\publisher}[2]{{\begin{center}\footnotesize\smalllineskip 
        Received #1\\
        Revised #2
        \end{center}
        }}

\def\abstracts#1#2#3{{
        \centering{\begin{minipage}{4.5in}\baselineskip=10pt\footnotesize
        \parindent=0pt #1\par 
        \parindent=15pt #2\par
        \parindent=15pt #3
        \end{minipage}}\par}} 

\newcommand{\bibit}{\nineit}

\renewenvironment{thebibliography}[1]
        {\frenchspacing
         \ninerm\baselineskip=11pt
         \begin{list}{\arabic{enumi}.}
        {\usecounter{enumi}\setlength{\parsep}{0pt}
         \setlength{\leftmargin 12.7pt}{\rightmargin 0pt} %FOR 1--9 ITEMS
         \setlength{\itemsep}{0pt} \settowidth
        {\labelwidth}{#1.}\sloppy}}{\end{list}}

\newcounter{itemlistc}
\newcounter{romanlistc}
\newcounter{alphlistc}
\newcounter{arabiclistc}

\newcommand{\fcaption}[1]{
        \refstepcounter{figure}
        \setbox\@tempboxa = \hbox{\footnotesize Fig.~\thefigure. #1}
        \ifdim \wd\@tempboxa > 5in
           {\begin{center}
        \parbox{5in}{\footnotesize\smalllineskip Fig.~\thefigure. #1}
            \end{center}}
        \else
             {\begin{center}
             {\footnotesize Fig.~\thefigure. #1}
              \end{center}}
        \fi}

\newcommand{\tcaption}[1]{
        \refstepcounter{table}
        \setbox\@tempboxa = \hbox{\footnotesize Table~\thetable. #1}
        \ifdim \wd\@tempboxa > 5in
           {\begin{center}
        \parbox{5in}{\footnotesize\smalllineskip Table~\thetable. #1}
            \end{center}}
        \else
             {\begin{center}
             {\footnotesize Table~\thetable. #1}
              \end{center}}
        \fi}

\def\@citex[#1]#2{\if@filesw\immediate\write\@auxout
        {\string\citation{#2}}\fi
\def\@citea{}\@cite{\@for\@citeb:=#2\do
        {\@citea\def\@citea{,}\@ifundefined
        {b@\@citeb}{{\bf ?}\@warning
        {Citation `\@citeb' on page \thepage \space undefined}}
        {\csname b@\@citeb\endcsname}}}{#1}}

\newif\if@cghi
\def\cite{\@cghitrue\@ifnextchar [{\@tempswatrue
        \@citex}{\@tempswafalse\@citex[]}}
\def\citelow{\@cghifalse\@ifnextchar [{\@tempswatrue
        \@citex}{\@tempswafalse\@citex[]}}
\def\@cite#1#2{{$\null^{#1}$\if@tempswa\typeout
        {IJCGA warning: optional citation argument 
        ignored: `#2'} \fi}}

\def\pmb#1{\setbox0=\hbox{#1}
        \kern-.025em\copy0\kern-\wd0
        \kern.05em\copy0\kern-\wd0
        \kern-.025em\raise.0433em\box0}

\def\fnt#1#2{\footnotetext{\kern-.3em
        {$^{\mbox{\scriptsize #1}}$}{#2}}}

\def\fpage#1{\begingroup
\voffset=.3in
\thispagestyle{empty}\begin{table}[b]\centerline{\footnotesize #1}
        \end{table}\endgroup}

\def\runninghead#1#2{\pagestyle{myheadings}
\markboth{{\protect\footnotesize\it{\quad #1}}\hfill}
{\hfill{\protect\footnotesize\it{#2\quad}}}}
\font\tenrm=cmr10
\font\tenit=cmti10 
\font\tenbf=cmbx10
\font\bfit=cmbxti10 at 10pt
\font\ninerm=cmr9
\font\nineit=cmti9

\font\eightrm=cmr8

\def\qed{\hbox{${\vcenter{\vbox{                        %HOLLOW SQUARE
   \hrule height 0.4pt\hbox{\vrule width 0.4pt height 6pt
   \kern5pt\vrule width 0.4pt}\hrule height 0.4pt}}}$}}

\renewcommand{\thefootnote}{\fnsymbol{footnote}}        %USE SYMBOLIC FOOTNOTE

\begin{document}

\runninghead{Spin-Flavour Symmetry and Contractions Towards 
Classical Space-Time Symmetry} {Spin-Flavour Symmetry and 
Contractions Towards Classical Space-Time Symmetry}

\normalsize\textlineskip
\thispagestyle{empty}
\setcounter{page}{1}

\copyrightheading{}                     %{Vol. 0, No. 0 (1993) 000--000}

\vspace*{0.88truein}

\fpage{1}
\centerline{\bf SPIN-FLAVOUR SYMMETRY AND CONTRACTIONS}
\vspace*{0.035truein}
\centerline{\bf TOWARDS CLASSICAL SPACE-TIME SYMMETRY}
\vspace*{0.37truein}
\centerline{\footnotesize ROLF DAHM\footnote{Permanent address: 
beratung f\"ur m.i.s., Herbststr. 24, D-99423 Weimar, Germany}}
\vspace*{0.015truein}
\centerline{\footnotesize\it Computing Center, University Mainz, 
rolf.dahm@uni-mainz.de}
\baselineskip=10pt
\centerline{\footnotesize\it D-55099 Mainz, Germany}
\vspace*{10pt}
\publisher{(received date)}{(revised date)}

\vspace*{0.21truein}
\abstracts{A classification scheme of hadrons is proposed on the basis 
of the division algebra {\bf H} of quaternions and an appropriate 
geometry. This scheme suggests strongly to understand flavour symmetry 
in another manner than from standard symmetry schemes. In our approach, 
we do {\it not} start from `exact' symmetry groups like SU(2)$\times$SU(2) 
chiral symmetry and impose various symmetry breaking mechanisms which 
collide with theorems wellknown from quantum field theory. On the contrary, 
the approximate symmetry properties of the hadron spectrum at low 
energies, usually classified by `appropriately' broken compact flavour 
groups, emerge very naturally as a low energy reduction of the noncompact 
(dynamical) symmetry group Sl(2,{\bf H}). This quaternionic approach not 
only avoids most of the wellknown conceptual problems of Chiral Dynamics 
but it also allows for a general treatment of relativistic flavour 
symmetries as well as it yields a direct connection towards classical 
relativistic symmetry.
}{}{}

\textheight=7.8truein
\setcounter{footnote}{0}
\renewcommand{\thefootnote}{\alph{footnote}}

\vspace*{1pt}\textlineskip      %) USE THIS MEASUREMENT WHEN THERE IS
\section{Introduction}          %) A SECTION HEADING
\label{intro}
\vspace*{-0.5pt}
\noindent
The standard approach towards a classification scheme of hadrons is still 
founded on Heisenberg's\cite{heisenberg} very old idea that the relevant 
flavour degrees of freedom can be described by a SU(2) symmetry group 
(`Isospin') and on Yukawa's\cite{yukawa} hypothesis of a (quantized) meson 
exchange to mediate nuclear forces. In the mean time, a lot of more 
sophisticated mechanisms have been added to this very basic concept to 
approach or even understand the plenty of available experimental data. 
However, to explain these data on the basis of SU(2) flavour or 
SU(2)$\times$SU(2) chiral theory almost all of the added ideas and mechanisms 
introduce further theoretical difficulties and sometimes even serious 
deficiencies. 

In the following, we'll start with a brief outline of symmetry methods where 
we focus especially on the manner of corrections applied to SU(2) isospin 
symmetry. Afterwards, we'll discuss a new symmetry approach which avoids 
these conceptual problems from the very beginning in that we do {\it not} 
start from a compact (flavour) symmetry group and try to relate it 
constructively to space-time symmetry. On the contrary, we understand compact 
flavour or chiral symmetry (which anyhow are appropriate descriptions only in 
the very low energy regime of the hadronic spectrum) as reasonable low energy 
approximation schemes of `the real' dynamical symmetry group Sl(2,{\bf H}). 
This noncompact symmetry group can be found by pure geometrical considerations 
as well as by trying to unify all the physical facts known from flavour, 
chiral and Wigner supermultiplet theory\cite{wigner}. We present a brief 
review of the resulting algebraic theory\cite{dahm2} \cite{dahm3} based on 
the Lie algebra isomorphism sl(2,{\bf H}) $\cong$ su$*$(4) $\cong$ so(5,1) 
before we focus on the route towards classical space-time symmetry.

\vspace*{1pt}\textlineskip      %) USE THIS MEASUREMENT WHEN THERE IS
\section{Flavour symmetry}      %) A SECTION HEADING
\label{flavour}
\vspace*{-0.5pt}
\noindent
At the beginning of the 60ies, field theoretical investigations of pions 
and pion-nucleon interactions\cite{bjoedre} at low energies on the basis 
of pure SU(2) isospin interactions led to results which were different 
from available data. Although the SU(2) pseudoscalar coupling scheme is 
suggested by a first sight on ratios of cross sections as well as by a 
naive comparison of SU(2) representations with the particle structure in 
the low energy regime of the hadron spectrum, it doesn't reproduce neither 
the pseudovectorial coupling of the pion to the nucleon nor the particle 
structure of the spectrum at higher energies nor transitions between SU(2) 
multiplets without introducing additional free parameters. Weinberg's 
pioneering work\cite{weinberg} \cite{dashen} on the pseudovectorial 
pion-nucleon coupling and more general investigations on effective 
Langrangians\cite{wesszum1} \cite{wesszum2} \cite{gasgeff} allowed to 
understand the pion coupling to the nucleon on a new footing with an 
underlying SU(2)$\times$SU(2) symmetry group, denoted by `Chiral Dynamics'. 
Although the larger chiral group respected the parity independence of 
strong interactions in addition to its charge independence (isospin) 
and although the chiral approach led to a reasonable description of the 
low-energy pion-nucleon coupling properties in terms of effective field 
theory, its strict interpretation has several serious deficiencies:
\begin{itemize}
\item The pseudovectorial coupling $\partial_{\mu}\vec{\pi}$ of the pion 
destroys renormalizability of the Lagrangian in perturbation theory and 
thus requires a {\it new} interpretation of Lagrangians in terms of 
`effective Lagrangians' as well as a redefinition of the applied methods 
termed to as `effective field theory'. 
\item In the hadron spectrum, SU(2)$\times$SU(2) representations of Chiral 
Dynamics are not realized via the Wigner-Weyl mode. To work around this 
defect, the concept of a spontaneously broken symmetry has been introduced 
to realize SU(2) isospin quantum numbers, and the three pion fields were 
interpreted as the necessary Goldstone bosons of this picture. However, 
Goldstone bosons have to be massless particles so that the observable mass 
of the pion triplet has to be restored by the further assumption of an 
additional explicit symmetry breaking mechanism, usually parametrized by 
partially conserved axial currents (PCAC theorem). Please note, however, 
that all these additional assumptions on symmetry breaking as well as on 
an identification of group representations are appropriate methods only 
when dealing with compact symmetry groups and a welldefined perturbation 
theory. In this case, all possible group actions connect only the finite 
number of states organized within the same irreducible representations of 
the group. A typical, very nice and simple example is quantum mechanics 
where the elements of the perturbation series
\begin{equation}
\label{eq:smatrix}
S_{fi}\,\sim\,
\left<\,f\,\right\|\,{\cal{H}}\,\left\|\,i\,\right>\,=\,
\left<\,f\,\right\|\,
\mbox{\it H}\,+\,\epsilon\,\mbox{\it H}\,'
\,\left\|\,i\,\right>
\end{equation}
can be understood well in terms of group theory. For compact groups, it 
is possible to use finite representation spaces according to properties 
of $|i>$ and $|f>$, to classify the symmetry breaking part $H'$ of the 
Hamiltonian ${\cal{H}}$ according to its covariance properties under 
symmetry transformations which leave the unperturbed Hamiltonian $H$ 
invariant, and to define appropriate measures on the representation spaces 
in order to calculate (\ref{eq:smatrix}). In addition, one may reparametrize 
$H$ in terms of an exponential of the Lie algebra, use theorems on 
completeness of matrix elements, etc. For compact groups, the concepts of 
spontaneously and explicitly broken symmetries can be understood within a 
simple geometrical framework\cite{dahm1}. However, in the context of a 
noncompact group or of SU(2)$\times$SU(2) Chiral Dynamics, a naive transfer 
of these symmetry concepts raises severe problems. Besides the fact that 
an appropriate representation theory as well as a suitable measure theory 
becomes very intricate or isn't even available, it has been shown for 
spontaneously broken symmetries that some of the symmetry generators 
(`charges') do not exist on the representation spaces of a (compact) 
subgroup as well as it makes no sense to argue about the magnitude of 
symmetry breaking\cite{fp1} \cite{fp2} \cite{jw}. Thus, dealing with 
SU(2)$\times$SU(2) Chiral Dynamics and identifying physical particles with 
SU$_{V}$(2) (isospin) representations, it is obvious that axial 
transformations mix different (irreducible) isospin representations 
(`superselection rules'), i.e. they mix the physical particle states of this 
picture (e.g. $\sigma$ $\leftrightarrow$ $\vec{\pi}$, $N$ $\leftrightarrow$ 
$\Delta$). Furthermore, it is {\it not possible} to represent the axial 
charges in terms of one-particle isospin states, and the discussion of a 
continuous limit $m_{\pi}^{2}$ $\to$ 0 of the symmetry breaking parameter 
has no theoretical foundation as a perturbative approximation of the 
welldefined `symmetry limit' $m_{\pi}^{2}=0$.

\item With respect to a relativistic symmetry formulation and the 
`No-Go-Theo\-rems'\cite{dashen}, it is interesting to note that Chiral 
Dynamics obviously suggests a nonstandard treatment of relating compact to 
noncompact (dynamical) symmetry groups. The symmetry scheme underlying 
Chiral Dynamics is far from being a supersymmetric one, and it also doesn't 
decouple compact and noncompact symmetry transformations via a direct product 
structure of the groups. On the contrary, a comparison of Chiral Dynamics 
with low energy data strongly suggests to couple the vector in isospin space 
with vectorial (p-wave) transformation properties under orbital angular 
momentum transformations\footnote{Using Weyl's unitary trick, it is 
straightforward to transfer the transformation properties of $\vec{L}$ with 
respect to homogeneous Lorentz transformations to the compact analogon 
\protect{SU(2)$\times$SU(2)/SU(2)} which is isomorphic to the symmetry 
scheme of Chiral Dynamics. Within the related noncompact and compact schemes, 
the diagonal subgroup \protect{SU$_{V}$(2)} may be identified with spatial 
rotational symmetry and with isospin, respectively, and the available 
low energy data strongly suggests to identify the pion fields in both schemes 
(isovector/pseudovector) within the adjoint representation of the diagonal 
subgroup. This behaviour is in excellent agreement with the viewpoint to 
understand flavour symmetry as a low energy approximation of an appropriate 
(noncompact) dynamical symmetry group.} $\vec{L}$. 
\end{itemize}
In the following, we present physical arguments as well as the algebraic 
foundation of our new ansatz which yields a lot of known symmetry properties 
of hadrons.

\vspace*{1pt}\textlineskip      %) USE THIS MEASUREMENT WHEN THERE IS
\section{Dynamic flavour symmetry}          %) A SECTION HEADING
\label{dynamics}
\vspace*{-0.5pt}
\noindent

\subsection{Physical motivation}
\noindent
Referring to Sudarshan's investigations\cite{sudarshan} on spontaneous 
symmetry breaking and approximate symmetries of hadrons, we have 
shown\cite{dahm2} \cite{dahm1} that various approaches to classify hadron 
multiplets in the low energy regime of the spectrum lead to an effective 
SU(4) theory. This effective symmetry scheme has the further advantage 
to yield a good description of hadron transformation properties, however, 
like in the case of Wigner's supermultiplet theory of nuclear ground 
states we are faced with an approximate (compact) symmetry group which 
yields a reasonable low energy description and describes a lot of observable 
symmetry properties but becomes worse at higher energies. This dynamical 
similarity in the behaviour of completely different physical systems 
motivates the viewpoint that the compact symmetry group SU(4) respectively 
its compact subgroups SU(2)$\times$SU(2), SU(2) and the structure 
SU(2)$\times$SU(2)/SU(2) known from flavour symmetry are low energy 
approximations of a suitable noncompact (dynamical) symmetry group. Using 
Weyl's unitary trick, we find SU$*$(4) which results from representing 
quaternions on complex vector spaces and which has the compact symmetry 
group Sp(2) $\cong$ USp(4) as maximal compact subgroup in common with SU(4).

\subsection{Mathematical motivation}
\noindent
A straightforward mathematical approach is guided by the necessity of a 
spinorial calculus to describe the observable half-integer spin/isospin 
representations suitably. The simplest case of spinorial calculus emerges 
in the framework of the stereographic projection $S^{2}$ $\to$ {\bf R}$^{2}$. 
To map all closed paths on $S^{2}$ (especially those passing through the 
north pole of the sphere!) appropriately into the plane 
it is necessary to introduce one hypercomplex unit $i$, i.e. to complexify 
the planar cartesian coordinates relative to each other. Thus, we may use 
either a complex planar variable $z$ and its complex conjugate or two real 
angles denoting points on the sphere. If we introduce in addition two 
homogeneous complex coordinates $z_{1}$ and $z_{2}$ related to $z$ via 
$z=z_{1}/z_{2}$, we can investigate M\"obius transformations of $z$,
\begin{equation}
\label{eq:moebius}
f\,:\,z\quad\to\quad f(z)\,=\,\frac{az+b}{cz+d}\,,\quad 
a,b,c,d\in\mbox{{\bf C}}\,,
\end{equation}
or an equivalent Sl(2,{\bf C}) matrix and spinor formalism which allows to 
apply the formalism of groups and appropriate representation theory. Thus, 
several different mathematical description are available for one and the 
same geometrical picture. 

An appropriate generalization of the spinorial concept to four dimensions, 
however, is {\it not} based on the group theoretical description. Instead,
eq.~(\ref{eq:moebius}) gives all the necessary hints if we study the 
projection $S^{4}$ $\to$ {\bf R}$^{4}$. To close paths through the `north 
pole', it is necessary to complexify the coordinates of ${\bf R}^{4}$ 
pairwise. Furthermore, in eq.~(\ref{eq:moebius}) multiplication as well as 
addition has to be defined between the `numbers', and inverse elements of the 
denominator have to exist. In order to choose the appropriate mathematical 
tools it is thus necessary to look at least for an algebra. Moreover, the 
function $f$ should have a {\it unique} singularity `to project the north pole 
to infinity'. This condition forbids zero divisors and thus restricts the 
possible choices from general algebras to the four division algebras with 
unit element. In the four dimensional case, we are thus left with Hamilton's 
quaternions. However, to avoid problems with quaternionic analysis we do not 
use generalized (quaternionic) M\"obius transformations $f(q)$ but the related 
matrix group Sl(2,{\bf H}) as well as SU$*$(4) and SO(5,1). 
The infinitesimal properties of these symmetry transformations are accessible 
via the Lie algebra sl(2,{\bf H}) and will be investigated in the next section 
on quaternionic, on complex and on real representation spaces due to the 
Lie algebra isomorphism sl(2,{\bf H}) $\cong$ su$*$(4) $\cong$ so(5,1). 
Vice versa, Lie theory allows to integrate the infinitesimal symmetry 
transformations.

\vspace*{1pt}\textlineskip      %) USE THIS MEASUREMENT WHEN THERE IS
\section{Algebraic approach to hadrons}          %) A SECTION HEADING
\label{algebra}
\vspace*{-0.5pt}
\noindent

\subsection{Sl(2,{\bf H}) and SU$*$(4)}
\noindent
The regular representation of Sl(2,{\bf H}) is isomorphic to the Dirac 
algebra\cite{dahm3} and thus provides the mathematical framework of quantum 
field theory whereas SU$*$(4) justifies to use SU(4) as an effective low 
energy approximation of the dynamical structure of the hadron 
spectrum\cite{dahm3} \cite{dahm1} \cite{dahm2}. SU(4) allows to identify 
nucleons and delta resonances within its spinorial third rank symmetric 
representation {\bf \underline{20}} and the mesons $\pi$, $\omega$ and 
$\rho$ within the spinorial second rank representation 
{\bf \underline{15}}\footnote{Screening the projective background of 
quaternionic spinors, the complex spinorial representations of mesons and 
nucleons/deltas suggest directly the `quark substructure' of hadrons, i.e. 
a threefold complex spinorial index for nucleon and delta degrees of freedom 
and a `dotted'/`undotted' pair of indices to denote the vectorial properties 
of the meson representation. Vice versa, a theory based on complex vector 
spaces {\it has} to reproduce a threefold spinorial index to describe hadronic 
fermions and a pair of conjugated indices to describe the mesons thus 
respecting the quaternionic foundations.}. The algebra used in Chiral 
Dynamics is completely contained within SU(4), and it is possible to discuss 
flavour and chiral transformation properties of hadrons in terms of SU(4) 
representations realized in the Wigner-Weyl mode\cite{dahm2}. It is noteworthy 
that the symmetry scheme Sl(2,{\bf H})/Sp(2) resp. its complex representation 
SU$*$(4)/USp(4) allows to identify F. Klein's ideas on complexified 
quaternions. Further investigations of Dirac theory in terms of complexified 
quaternions show that the mass is an arbitrary parameter which drops out 
completely, and that we are left with a theory using velocities as basic 
parameters\cite{dahm1} \cite{dahm3}. This leads to a Lobachevskian geometry 
and a relation between parameters of the Lie algebra sl(2,{\bf H}) (or 
su$*$(4)) and {\it velocities}\cite{dahm3} (see also\cite{smorodinskii}).

\subsection{SO(5,1) and Classical Space-Time}
\noindent
The Lie algebra so(5,1) contains the de Sitter algebra so(4,1) which can 
be Wigner-In\"{o}n\"{u}-contracted\cite{wigino} towards the Poincar\'{e} 
algebra and to the Galilei algebra in the limits of vanishing curvature 
(R $\to$ $\infty$) of the universe and vanishing ratio $v/c$ ($c$ $\to$ 
$\infty$), respectively\cite{gilmore}. Here, we don't review all the 
technical details but it is noteworthy that none but the algebras so(4,1) 
and so(3,2) may be contracted to the Poincar\'{e} algebra\cite{levynahas}. 
It is a second important property of contractions that widely used 
differential operators like $\partial_{\mu}$ appear only {\it after} 
contraction of the generators of Lie groups. Appropriately, this 
description is isomorphic to projective (`flat') physics where the 
operators $\partial_{\mu}$ represent contracted elements of the full Lie
algebra so(4,1) which itself spans a vector space, the tangent space to 
the Lie group SO(4,1) at unity. As a direct consequence, the definition 
of observable `mass' is related {\it only} to the contracted limit of 
the more general Lie algebra so(4,1) so that there is no foundation to 
introduce mass parameters in the framework of group transformations acting 
on homogeneous coordinates. This suggests to understand mass as a classical 
effective parameter which emerges after the contraction process, and it 
explains why the mass parameters drop out when using complexified 
quaternions to describe Dirac theory. In addition, if we interpret 
interactions in Dirac theory on the basis of sl(2,{\bf H}) transformations 
acting on homogeneous quaternionic coordinates, these interactions 
necessarily change the `masses' of the involved particles.

\end {document}